# Multivariate brain-cognition associations in euthymic bipolar disorder

## AUTHORS

Bethany Little*[1,2], Carly Flowers[1], Andrew Blamire[1], Peter Thelwall[1], John-Paul Taylor[1], Peter Gallagher[1], David Andrew Cousins†[1,3], Yujiang Wang†[1,2]

## AFFILIATIONS

[1]Translational and Clinical Research Institute, Faculty of Medical Sciences, Newcastle University, Newcastle upon Tyne, United Kingdom

[2]CNNP Lab (www.cnnp-lab.com), Interdisciplinary Computing and Complex BioSystems Group, School of Computing, Newcastle University, Newcastle upon Tyne, United Kingdom

[3]Cumbria, Northumberland, Tyne and Wear NHS Foundation Trust, Newcastle upon Tyne, United Kingdom

†Joint last authors

*Corresponding Author: bethany.little@ncl.ac.uk, CNNP Lab, Interdisciplinary Computing and Complex BioSystems Group, School of Computing, Newcastle University, Urban Sciences Building, 1 Science Square, Newcastle upon Tyne, United Kingdom, NE4 5TG.

## ACKNOWLEDGEMENTS

We thank the participants of the Bipolar Lithium Imaging and Spectroscopy Study (BLISS) study for contributing their time. The BLISS project was funded by the Medical Research Council (Clinician Scientist Fellowship BH135495 to DAC). BL was supported by the Leverhulme Doctoral Scholarship in Behaviour Informatics (DS-2017-015). YW was supported by a UKRI Future Leaders Fellowships (MR/V026569/1). BL and YW were supported by the EPSRC (EP/Y016009/1). The funders did not have a role in study conception, design, data collection, data analysis, interpretation of the data, preparation of the article, or article submission.

## CONFLICT OF INTEREST

None.

# ABSTRACT

**Background**: People with bipolar disorder (BD) tend to show widespread cognitive impairment compared to healthy controls. Impairments in processing speed (PS), attention, and executive function (EF) may represent 'core' impairments that have a role in wider cognitive dysfunction. Cognitive impairments appear to relate to structural brain abnormalities in BD, but whether core deficits are related to particular brain regions is unclear and much of the research on brain-cognition associations is limited by univariate analysis and small samples.

**Methods**: Euthymic BD patients ($n$=56) and matched healthy controls ($n$=26) underwent T1-weighted MRI scans and completed neuropsychological tests of PS, attention, and EF. We utilised public datasets to develop a normative model of cortical thickness ($n$=5,977) to generate robust estimations of cortical abnormalities in patients. Canonical correlation analysis was used to assess multivariate brain-cognition associations in BD, controlling for age, sex, and premorbid IQ.

**Results**: BD showed impairments on tests of PS, attention, and EF, and abnormal cortical thickness in several brain regions compared to healthy controls. Impairments in tests of PS and EF were most strongly associated with cortical thickness in left inferior temporal, right entorhinal, and right temporal pole areas.

**Conclusion**: Impairments in PS, attention, and EF can be observed in euthymic BD and may be related to abnormal cortical thickness in temporal regions. Future research should continue to leverage multivariate methods to examine complex brain-cognition associations in BD. Future research may benefit from exploring covariance between traditional brain structural morphological metrics such as cortical thickness, cortical volume, and surface area.

# MeSH Keywords

# INTRODUCTION

People with bipolar disorder (BD) tend to show widespread cognitive impairment compared to healthy controls.[1,2] BD is also associated with structural brain abnormalities, including abnormal cortical thickness, compared to healthy controls across widespread areas of the brain.[3] Both cognitive impairments and brain abnormalities can be observed in euthymia, suggesting they represent core features or consequences of BD.[2,4] Some studies suggest a relationship between brain structure and cognitive impairment in BD,[4-7] however there is currently a lack of multivariate studies.

Processing speed (PS), attention, and executive function (EF) are often considered primary or 'core' cognitive impairments in BD and may explain wider cognitive dysfunction, such as poor memory.[8-10] From this perspective, their potential relationship to abnormal brain morphology may be particularly important. Some research suggests that performance on tests of PS, attention and EF may indeed relate to cortical thickness in BD, even in euthymia.[5,11] Neuropsychological tests that capture these processes, such as the Digit Symbols Substitution Test (DSST) and Trail-Making Test (TMT), often detect a large magnitude of impairment[12] and have been specifically related to cortical thickness of brain regions in BD.[11] However, findings are inconsistent with not all studies detecting associations.[13,14] Few previous studies have focussed on the aforementioned cognitive processes, and studies often omit tests of PS and attention or focus instead on general cognitive functioning. Thus, the complex nature of the relationship between core cognitive dysfunction and cortical thickness is not well understood and there is no consensus on the specific brain regions involved.

Several challenges currently limit brain-cognition studies: recent research suggests that brain-behaviour associations are not replicable,[15] and results appear inconsistent with some studies failing to detect significant structural brain-cognition associations.[14,16,17] This may be driven by small sample sizes and previous research generally relies on univariate analysis, where the relationship of a single cognitive function and the structure of several brain regions is tested. Studying cognitive functions independently may not be an accurate model of cognitive impairment in BD, since processes likely interact and overlap, especially in the presence of cognitive impairment.[8,9,18] The univariate approach is also constrained by the problem of multiple comparisons, where there is a risk of missing real associations after correcting for multiple comparisons (i.e., increased type-II errors). Multivariate methods that can simultaneously accommodate the multivariate cognitive and brain data may better reflect the complex, multimodal nature of behaviour and physiology. Such methods are beginning to be leveraged to assess brain-cognition relationships,[5,19,20] however, few studies have utilised these methods, especially in BD.

Here, we examined multivariate associations between cortical thickness and core cognitive functions in euthymic BD. Further, we used large public datasets to build a normative model of cortical thickness to produce more robust estimations of structural abnormalities in BD.[21] We hypothesised that abnormalities in cortical thickness in groups of regions would be associated with poor core cognitive performance, particularly in PS and attention.

## METHODS

Data were collected as part of the Bipolar Lithium Imaging and Spectroscopy Study (BLISS).[22] T1-weighted imaging, diffusion-weighted imaging and lithium imaging data have been reported previously.[22,23] BLISS was granted a favourable ethical opinion by a United Kingdom National Research Ethics Committee (14/NE/1135). All participants provided written informed consent.

## Participants

Fifty-nine patients with BD were enrolled in the study had neuropsychological and imaging data suitable for analysis. Inclusion criteria for patients were a diagnosis of BD (type I or II) according to DSM-5 criteria. Diagnosis was confirmed through clinical interview by an experienced research assistant (CF), supervised by a senior psychiatrist (DAC) via discussion of the assessment and suitability for inclusion. Comorbid psychiatric diagnoses were permissible (excluding neurodevelopmental, neurocognitive, neurodegenerative, and substance abuse disorders) if the primary diagnosis was BD, confirmed by a senior psychiatrist (DAC) reviewing case notes as required. All patients were euthymic at study entry and at assessment, defined as a score of <7 on both the Young Mania Rating Scale (YMRS)[24] and the 21-item Hamilton Rating Scale for Depression (HAM-D).[25] Twenty-nine BD patients were taking lithium; the other patients were taking maintenance treatments but were naïve to lithium. Twenty-eight matched healthy controls (HC) were recruited, all of whom had no history of psychiatric illness and were not taking any psychotropic medications. All subjects were 18-65 years of age. Exclusion criteria for all subjects were: contraindications to MR examination (e.g., cardiac pacemaker claustrophobia, exceeding 150kg in weight), current or past medical condition likely to affect brain structure; current or recent substance abuse (NetSCID Module E); alcohol intake exceeding 21 units per week (self-reported); and a learning disability or impairment of capacity.

## Materials

### Clinical measures

The NetSCID diagnostic tool was used to confirm diagnosis and assess psychiatric comorbidities; this is a validated online version of the Structured Clinical Interview for DSM-V criteria (Telesage, Inc.,

Chapel Hill, NC, USA). YMRS and HAM-D measured current manic and depressive symptoms, respectively.

### Neuropsychological tests

A battery of pen-and-paper neuropsychological tests[26] was used to measure PS, attention, and EF as follows.

The d2 test measured attention and concentration. Participants crossed out as many target stimuli as possible (the letter 'd' with two dashes above or below it) amongst a series of distractor stimuli (letters 'd' or 'p' with any other number of dashes around it) within 20 seconds. Three outcome measures were recorded: the percentage of correct responses; the total number of correctly cancelled minus total number incorrectly cancelled (concentration performance); and the maximum minus the minimum of total items processed in a trial (fluctuation rate).

The DSST measured PS and attention. Participants transcribed unique geometric symbols with corresponding Arabic numbers using a key on the page. Participants also completed a sub-test of the DSST, Symbol Copy, which involves copying symbols from one row of boxes to another, retaining elements of motor speed but not requiring other cognitive processes involved in transcribing. We subtracted the time taken to complete the Symbol Copy version from the time taken to complete the DSST original version to reflect cognitive PS without the motor component.

Part A of the TMT was used to measure PS and attention. Participants connected a series of numbered circles spread on a page in ascending order. Part B of the TMT tested EF. Participants connected a series of circles that either contained a number or letter, in alternate ascending and alphabetical order (e.g., 1-A-2-B-3-C, etc.). In both parts, the time to complete the task was measured. TMT-B tests working memory and switching (elements of EF), as well as motor speed, cognitive speed, and attention. The latter three functions were fractioned out of the TMT-B score by subtracting the time to complete TMT-A from the time to complete TMT-B, leaving a measure of EF. Category fluency was also used to measure EF, where participants named as many animals as possible in 60 seconds. The number of correctly named animals was recorded.

# Procedure

All subjects attended a screening visit to confirm eligibility and complete structured clinical interview.

### Scanner and image acquisition

MRI scans were performed at 9am. MR data were acquired on a 3T Philips Achieva MRI scanner (Philips Medical Systems) with $^1$H structural and diffusion weighted imaging performed using a

Philips 8-channel SENSE head coil. For each subject, we acquired 3D $T_1$-weighted images (T1w) of brain anatomy with a $^1$H gradient echo sequence (TR=9.6ms, TE=4.6ms, FOV=240×240×180mm, acquisition matrix=240x208x180mm, voxel size=1x1.15x1mm, reconstructed into matrix size=256x256x180mm, 1mm$^3$ average), and 2D $T_2$-weighted Fluid-Attenuated Inversion Recovery (FLAIR) images (50 slices, TR=11000ms, TE=125ms, FOV=240×240mm, matrix size=256x256mm, voxel size=1x1x3mm average).

## Data analysis

### MRI data pre-processing

All BLISS images were exported in DICOM format and converted to NIFTI format using the *dcm2niix* program (v1.0.20180622). Data pre-processing and analysis were performed using FreeSurfer 6.0,[27] utilising the standard *recon-all* pipeline, including removal of non-brain tissue, segmentation of grey matter and white matter surfaces, and cortical parcellation. FLAIR images were used in the *recon-all* pipeline to improve the pial surface and segmentation. Subjects were excluded if the T1w images included MR artefacts or were poor quality (*n*=1; HC). For the remaining subjects, visual quality inspections were performed on a sub-sample of participants (approximately half the total sample; *n*=30 BD and *n*=18 HC) in FreeSurfer's FreeView function. Manual corrections of the pial surface were conducted where necessary using FreeView (*n*=24 subjects were edited). After manual corrections, the overall image quality and FreeSurfer processing was deemed adequate and cortical thickness measures were obtained for 68 anatomical regions of interest (ROIs) for each subject using the Desikan-Killiany atlas.

### Normative model of cortical thickness

To mitigate issues with small matched healthy control samples, we developed a normative model of cortical thickness using public datasets of healthy controls.[28-31] A sample of *n*=6,344 healthy controls were included (normative-HC: mean age 32.64; 50.3% females; see supplementary materials for further details). All data were pre-processed using the standard *recon-all* pipeline in FreeSurfer. We harmonised the normative-HC data, including the 26 HC subjects recruited to BLISS, for each ROI by regressing age, sex and scanning protocol (i.e., site; as a random effect) on cortical thickness using a generalised additive model. The correction was performed in R using the package *mgcv* (https://CRAN.R-project.org/web/packages/mgcv). We removed controls who had outliers in any ROI, defined as residuals greater than 5 mean absolute deviations from the mean (*n*=393).

The BD patient data were then adjusted for age, sex, and site, using the prediction from the normative model, where the site correction was informed by the site correction of the BLISS HC subjects. Data for each ROI were then z-scored based on the control group mean and standard deviation. The resulting data therefore reflected abnormalities in cortical thickness for each ROI for each patient, based on a robust normative model of cortical thickness, accounting for age, sex, and scanning protocol.

## Neuropsychological data pre-processing

Missing values for pre-morbid IQ (NART) and neuropsychological variables in the BLISS study were imputed using linear regression (5% of data; see supplementary materials for details). Variables were reversed where appropriate so that a higher score reflects better performance. Age and pre-morbid IQ were regressed out of the neuropsychological data in the HC group using robust regression, then this model was used to regress age and pre-morbid IQ out of the BD data. Standardised residuals following age and pre-morbid IQ regression were standardised (z-scored) based on the HC group mean and standard deviation so that the neuropsychological data reflected an abnormality from the matched controls.

## Statistical analysis

Statistical analysis was performed using R Studio version 4.1.2.[32] Tests were regarded as significant if $p<.05$. Data were inspected to ensure assumptions of each statistical model were met and neuropsychological data were transformed as required to fit assumptions of multivariate normality (see supplementary materials). The subsequent analysis was run using the untransformed neuropsychological data as well to compare results; those results are presented in the supplementary materials.

### *Group differences*

Student's t-tests were used to compare groups where data were normally distributed, otherwise non-parametric Mann-Whitney U-tests were used. Group differences in dichotomous variables were analysed using $\chi 2$ tests. Cohen's *d* was used to estimate effect sizes. Group differences in cortical thickness for each ROI were corrected for multiple comparisons using the Benjamini-Hochberg method to control the false discovery rate, which was set at 5%.[33]

### *Canonical Correlation Analysis (CCA)*

Canonical Correlation Analysis (CCA) was used to assess multivariate associations between neuropsychological performance and cortical thickness of regions in the BD group. CCA reduces the number of variables in each set to latent variables (canonical variates) that maximise the correlation between two sets of data. The output of CCA is correlated pairs of canonical variates, which are ordered with the first canonical pair (i.e., *U*1 and *V*1) explaining the largest proportion of covariance between the two sets. Canonical loadings represent the relationship between an original variable in the dataset and the canonical variate of its own dataset, and canonical *cross*-loadings represent the relationship between an original variable in the dataset and the canonical variate of the other dataset. Loadings and cross-loadings with a coefficient above 0.3 or below -0.3 were considered non-trivial.[19]

Since the number of BD subjects was less than the number of ROIs, a Principal Components Analysis (PCA) was applied to the cortical thickness dataset before CCA to reduce dimensionality to limit the feature-to-sample ratio and avoid overfitting.[19,34] The number of principal components (PCs) that collectively explained 90% of this dataset were retained and fed into the CCA as the cortical thickness dataset ($V$). Cross-loadings for each original variable (e.g., ROIs) were estimated by correlating the original data for each ROI with the canonical variate $U1$.

The statistical significance of the canonical correlations was tested using Pillai's trace as a conservative multivariate test of significance.[35] We also ran a permutation test, where subjects were randomly shuffled in the cortical thickness dataset ($V$), then CCA was performed on the mis-matched datasets. This was done 10,000 times to yield a distribution of statistics produced with randomly shuffled data. We would expect a true significant result of the original CCA to have a more extreme value than the results from the permuted data, allowing for 5% of extreme values to be produced by chance. The percentage of randomly permuted results that were more extreme than the original CCA can be interpreted as a $p$-value, where $p<.05$ suggests statistical significance.

# RESULTS

## Group differences in demographic, clinical and neuropsychological data

Table 1 shows demographic information and clinical characteristics for patients with BD and HC subjects. Groups did not differ in age or sex. HC had significantly more years of education than BD. BD scored significantly higher on HAM-D and YMRS than HC but were within the range consistent with euthymic mood. Groups did not differ in premorbid IQ (NART score) or on average cortical thickness in each hemisphere. Table 1 also displays group differences in raw neuropsychological scores: BD patients showed significantly poorer performance than HC subjects on most tests, with medium effect sizes ($d>0.5$). Figure 1 illustrates neuropsychological group differences after correcting for age and premorbid IQ and imputing missing values. The results showed a similar pattern of deficits in BD, except TMT-B (minus TMT-A) was not significant ($p=.063$, $d=0.51$; see supplementary materials).

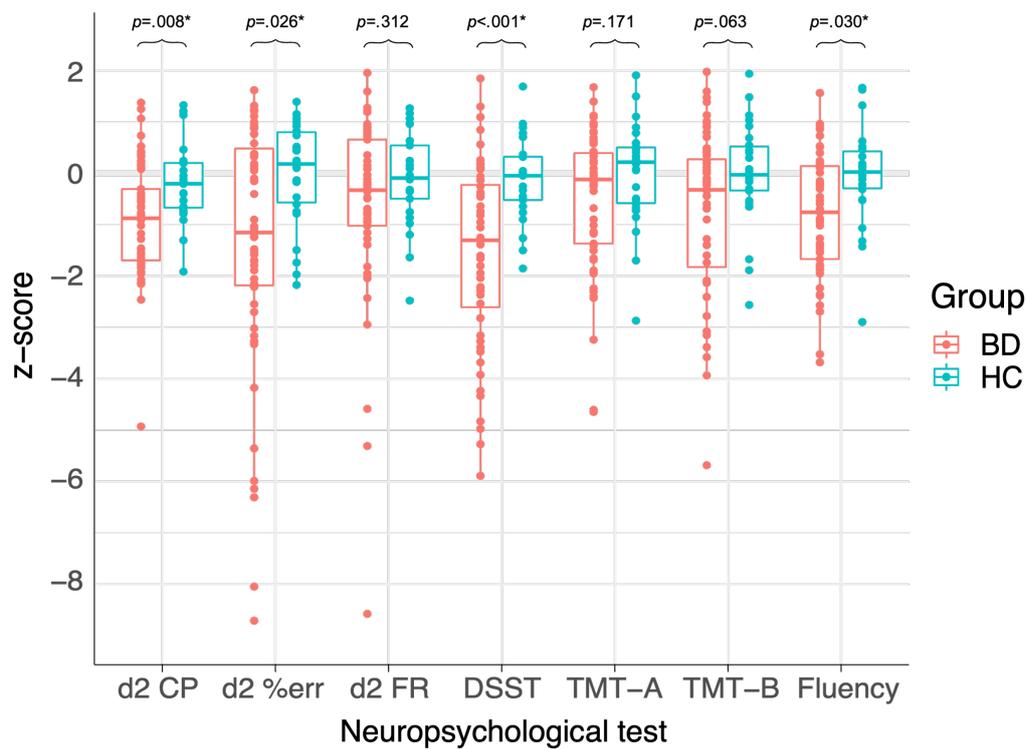

Figure 1: Neuropsychological performance for bipolar disorder patients (BD; *n*=56) and matched healthy control (HC; *n*=26) group. The data presented here are the pre-processed data where missing data were imputed, age and premorbid IQ was regressed out, data were z-scored based on the control group mean and standard deviation, and reversed where appropriate so that lower z-scores reflected worse performance. DSST score was calculated by subtracting the Symbol Copy score from the original Digit Symbol Substitution score, and TMT-B score was calculated by subtracting TMT-A time from TMT-B time. CP=concentration performance; %err=percentage of errors; FR=fluctuation rate, DSST=Digit Symbol Substitution Test; TMT=Trail-Making Test.

## Cortical thickness in BD patients

Figure 2 shows group differences in cortical thickness for each ROI between the BD group and the normative model. BD showed thinner cortex than normative-HC in most areas; group means of cortical thickness for each ROI and associated group differences are provided in the supplementary materials. Medium effect sizes were found for left postcentral, precentral, paracentral, and caudal middle frontal areas (*d*>.5). Small effect sizes were found for left precuneus and cuneus; and right paracentral, precentral, postcentral, insula, and pericalcarine areas (*d*>.2). As an exploratory step, we tested whether cortical thickness differed between patients taking lithium and patients not taking lithium in our sample; no significant group differences were found after correcting for multiple comparisons (all *ps*>.05).

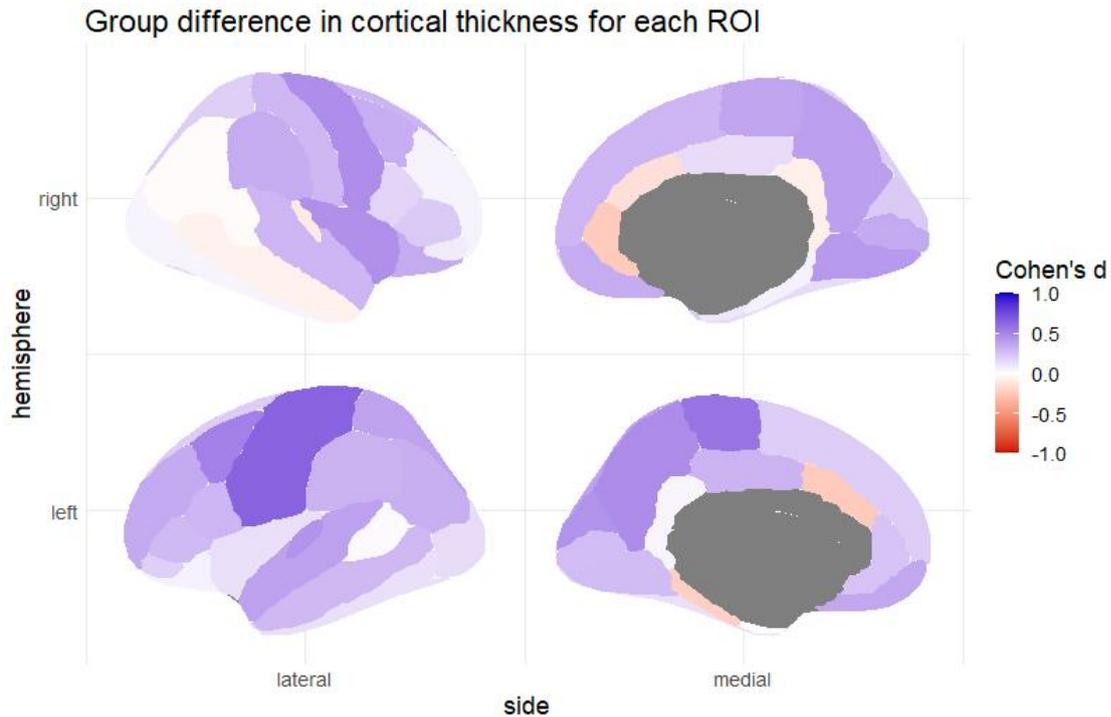

Figure 2: Effect size (Cohen's *d*) of the group differences between bipolar disorder patients (*n*=56) and healthy controls from the normative model (*n*=5,977) in cortical thickness for each region of interest (ROI).

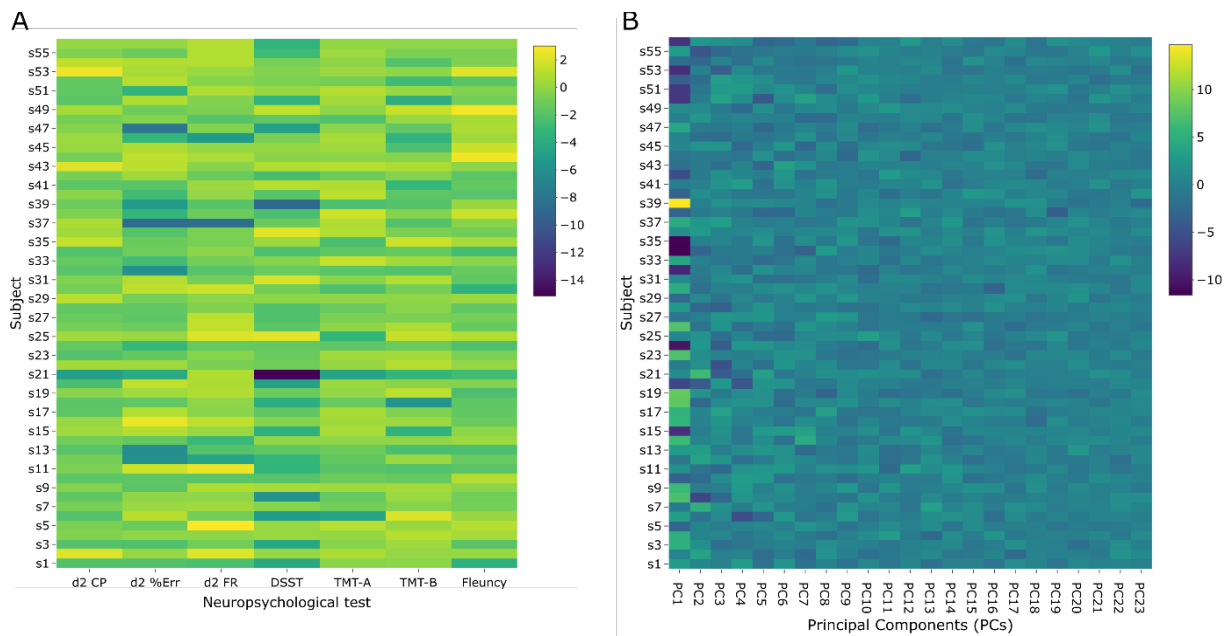

Figure 3: Heatmaps of neuropsychological and cortical thickness data for each bipolar disorder (BD) subject. A) Neuropsychological data for each outcome measure z-scored based on control group mean and standard deviation (dataset *U*). B) Cortical thickness principal components (PCs); dataset *V*). CP=concentration performance; %Err=percentage of errors; FR=fluctuation rate; DSST=Digit Symbol Substitution Test; TMT-A=Trail-Making Test.

|  | Healthy Control group | | | | Bipolar Disorder group | | | | Group differences | | | |
| --- | --- | --- | --- | --- | --- | --- | --- | --- | --- | --- | --- | --- |
|  | n | Mean | SD | Median | n | Mean | SD | Median | Statistic (t/W) | df | p | Cohen's d |
| **Demographics and clinical characteristics** | | | | | | | | | | | | |
| **Sex (n female, % female)** | 26 | n=12 | 46% | - | 56 | n=35 | 62% | - | 1.33 | 1 | .249 | - |
| Age (years) | 26 | 48.46 | 11.80 | 48.50 | 56 | 45.36 | 12.15 | 46.00 | 612.5 | - | .252 | 0.26 |
| Education (years) | 26 | 16.46 | 3.78 | 16.00 | 56 | 14.14 | 2.88 | 13.00 | 470.0* | - | .009 | 0.74 |
| HAM-D | 26 | 1.00 | 1.57 | 0 | 56 | 4.20 | 4.16 | 3.00 | 1123.5* | - | <.001 | 0.91 |
| YMRS | 26 | 0.12 | 0.43 | 0 | 56 | 1.06 | 2.23 | 0 | 938.0* | - | .007 | 0.51 |
| **Medication (n lithium, % lithium)** | - | - | - | - | 56 | n=29 | 52% | - | - | - | - | - |
| **Neuropsychological data** | | | | | | | | | | | | |
| NART IQ | 24 | 15.54 | 7.54 | 14.50 | 52 | 17.48 | 6.76 | 18.00 | 1.08 | 40.70 | .288 | 0.28 |
| d2 Concentration Performance | 25 | 177.92 | 33.41 | 177.00 | 55 | 154.15 | 45.12 | 144.00 | 439.5* | - | .010 | 0.58 |
| d2 Percent Error | 25 | 0.03 | 0.02 | 0.03 | 55 | 0.05 | 0.05 | 0.04 | 876.5 | - | .050 | 0.57 |
| d2 Fluctuation Rate | 25 | 10.64 | 2.78 | 11.00 | 55 | 12.11 | 4.99 | 12.00 | 809.5 | - | .205 | 0.34 |
| DSST Symbol (minus Symbol Copy) | 23 | 73.12 | 13.28 | 74.50 | 53 | 93.03 | 35.33 | 88.55 | 856.0* | - | .005 | 0.66 |
| TMT-A time | 23 | 27.71 | 11.90 | 26.55 | 54 | 30.64 | 11.31 | 28.37 | 725.0 | - | .249 | 0.26 |
| TMT-B (minus TMT-A) | 23 | 27.52 | 16.96 | 23.40 | 52 | 39.16 | 24.02 | 32.09 | 770.0* | - | .049 | 0.53 |
| Category Fluency total | 23 | 24.96 | 4.96 | 25.00 | 54 | 21.80 | 6.24 | 21.50 | -2.36* | 51.85 | .022 | 0.54 |
| **Cortical thickness** | | | | | | | | | | | | |
| Average left hemisphere | 26 | 2.46 | 0.10 | 2.45 | 56 | 2.43 | 0.09 | 2.42 | -1.70 | 46.82 | .095 | 0.42 |
| Average right hemisphere | 26 | 2.46 | 0.09 | 2.45 | 56 | 2.43 | 0.10 | 2.41 | -1.28 | 50.23 | .206 | 0.30 |

Table 1: Descriptive statistics and group differences on demographics, clinical characteristics, raw neuropsychological variables (i.e., data before pre-processing), and average cortical thickness, for BD patients (*n*=56) and the matched healthy control group (*n*=26). *Significant at *p*<.05. SD=standard deviation; *df*=degrees of freedom; HAM-D= Hamilton Rating Scale for Depression; YMRS=Young Mania Rating Scale; NART=National Adult Reading Test; DSST=Digit Symbol Substitution Test; TMT=Trail-Making Test.

# Canonical Correlation Analysis (CCA)

PCA was conducted on the cortical thickness data: 23 PCs explained 90% of the data. CCA was performed using the seven neuropsychological variables as dataset *U* and the 23 cortical thickness PCs as dataset *V*; Figure 3 shows heatmaps of each dataset. Figure 4 shows the first canonical correlation coefficient, which had a strong linear correlation of .860. Pillai's Trace suggested that this was not statistically significant (*p*=.112; see supplementary materials for results for the remaining canonical correlations). A permutation test with *n*=10,000 permutations also suggested the correlation was not significant but was trending towards significance (*p*=.093).

Table 2 displays the canonical loadings cross-loadings for each cognitive variable from dataset *U*. DSST (minus Symbol Copy) and TMT-B (minus TMT-A) were most strongly associated with the first cortical thickness canonical variate *V1*. Figure 5 illustrates the association between cortical thickness of each region in dataset *V* and the cognitive dataset (*U*). Three regions were most strongly associated with the first cognitive canonical variate *U1*: left inferior temporal (*rho*=.406, *p*=.002), right entorhinal (*rho*=.380, *p*=.004) and right temporal pole areas (*rho*=.334, *p*=.012). Three other regions did not reach the threshold of .3 but had significant *p*-values: right inferior temporal (*rho*=.286, *p*=.032), right pars orbitalis (*rho*=.268, *p*=.046), and right superior temporal (*rho*=.297, *p*=.026) areas. The full list of cross-loadings is provided in the supplementary materials.

| Variables from dataset *U* | Loadings | Cross-loadings |
|---|---|---|
| d2 Continuous Performance | .233 | .201 |
| d2 Percentage of Errors | .222 | .191 |
| d2 Fluctuation Rate | -.082 | -.070 |
| DSST (minus Symbol Copy) | -.675 | -.580* |
| TMT-A | -.118 | -.101 |
| TMT-B (minus TMT-A) | .353 | .303* |
| Category Fluency | .114 | .098 |

Table 2: Canonical loadings and cross-loadings for each variable in the cognitive dataset (*U*). *Associated with the first canonical variate *V1* at the 0.3 threshold. The neuropsychological data used in the CCA were the pre-processed data where missing data were imputed, age and premorbid IQ was regressed out, data were z-scored based on the control group mean and standard deviation, and reversed where appropriate so that lower z-scores reflected worse performance. DSST score was calculated by subtracting the Symbol Copy score from the original Digit Symbol Substitution score, and TMT-B score was calculated by subtracting TMT-A time from TMT-B time. DSST=Digit Symbol Substitution Test; TMT=Trail-Making Test.

Results of the CCA for the *untransformed* neuropsychological data are provided in the supplementary materials (*i.e.*, pre-processed data with missing data imputed, age and NART IQ regressed out, and z-scored, but not transformed to address skewed distributions). The overall result was similar, with the first canonical pair showing a strong canonical correlation that did not reach significance. The pattern of cross-loadings differed slightly, with d2 concentration performance, DSST and TMT-A showing the strongest cross-loadings (all <-.3). The regions with the strongest cross-loadings were also slightly different, with the left insula, and right entorhinal, inferior temporal and medial orbitofrontal reaching the .3 threshold.

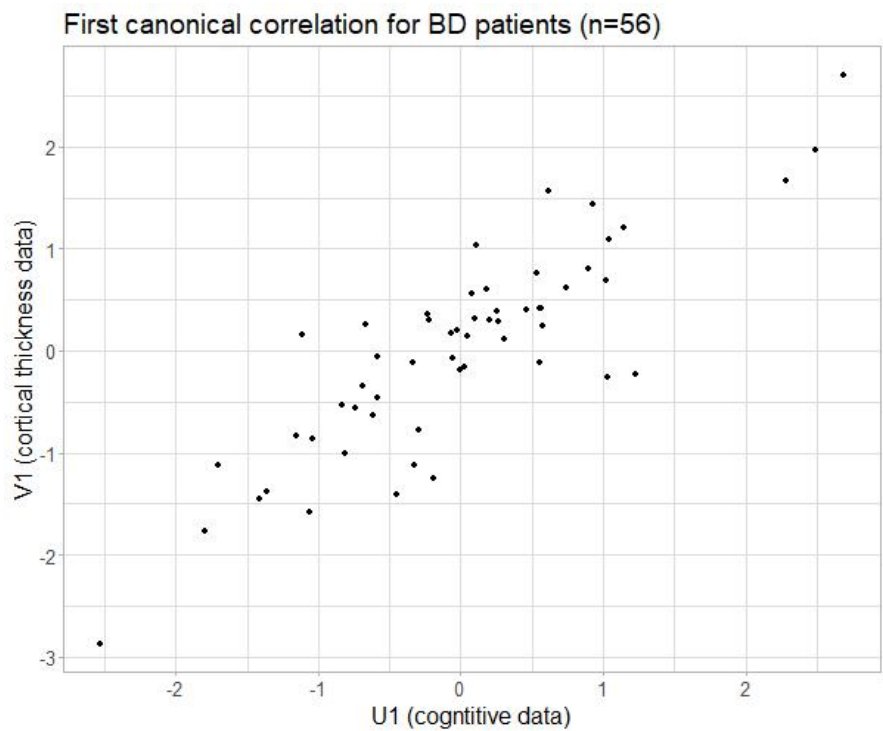

**Figure 4: Scatterplot of the first canonical correlation for bipolar disorder (BD) patients (*n*=56).**

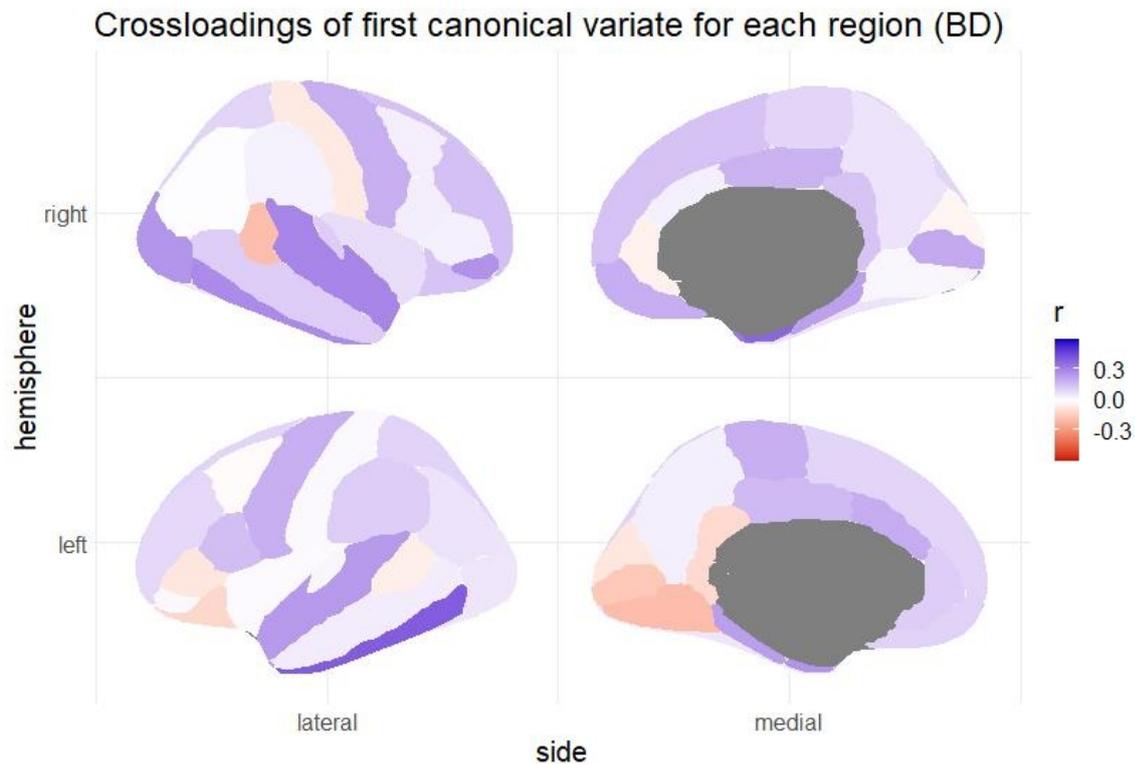

Figure 5: Canonical cross-loadings illustrating the strength of association between cortical thickness in each region and the first canonical variate from the cognitive data (*U1*).

# DISCUSSION

We examined multivariate associations between cognitive impairments and abnormal cortical thickness in euthymic BD. The BD group showed poorer performance than matched HC group on some tests of PS, EF, and attention, even after accounting for age and premorbid IQ. Multivariate cross-loadings suggested that impairments in processing speed (DSST) and executive function (TMT-B) were associated with abnormal cortical thickness in left inferior temporal, right entorhinal, and right temporal pole areas in our BD sample. However, the canonical correlation did not reach statistical significance, most likely due to the limited patient sample size in our study.

Our results are in line with previous research showing cortical thinning in BD in bilateral frontal and parietal regions.[3,36] Our results are also consistent with studies showing impairments in core cognitive functions in BD.[8–10,37] We did not find an impairment in TMT-A in our BD sample, whereas previous research suggested that TMT-A is one of the most robust measures of cognitive impairment in euthymic BD.[37] This discrepancy may be explained by our sample size and sources of heterogeneity that we did not control for here, e.g., illness severity and psychosis.[2] Alternatively, our null results may be a consequence of regressing out age and premorbid IQ. Interestingly, while there was no group difference in TMT-B (minus TMT-A), abnormal scores in this test appeared to be related to abnormal cortical thickness. Other studies have similarly found correlations between TMT and cortical thickness in euthymic BD where there was no group difference between BD and healthy controls;[36] these findings may suggest that comparing group means may not be sensitive to some of the cognitive abnormalities found in BD that tend to be heterogeneous within groups.

While previous studies found associations between cognitive ability and brain structure in mostly frontal and parietal regions in BD,[5,6] the most strongly associated areas in our study were in the temporal lobe. Associations between temporal regions and cognitive function have been observed in BD,[6] including associations between working memory and right temporal pole being found specifically in BD but schizophrenia patients.[7] Other studies have shown that PS (measured with DSST)[11] and executive function[36] were related to cortical thickness specifically in temporal and parietal regions, including in euthymia.[36] Our finding that the entorhinal area was associated with cognitive dysfunction may be in line with research showing associations between structural abnormalities in fronto-limbic areas and poor cognitive functioning in BD.[38] However, comparing our results to previous literature should be done with caution, given that our results took an explicitly multivariate approach. Nevertheless, our study demonstrates the utility of normative models and multivariate techniques to assess complex brain-cognition association in mood disorder groups.

## Limitations

This study has several limitations that should be considered: firstly, the neuropsychological test battery included measures of core cognitive functions but did not test other cognitive functions such as working memory, verbal memory, or visuo-spatial processing. Thus, we did not assess the entire spectrum of potential brain-cognition associations. The tests used here were pen-and-paper tests; computerised tests of reaction time may provide more accurate measures of PS and attention. Similarly, the d2 measure of attention does not specifically capture sustained attention, which is impaired in euthymic BD,[39] first-degree relatives of patients,[40] and may reflect a core feature of the disorder.[10] Sustained attention has been found to be the most strongly related to mood disorders out of other attentional processes.[41] Our results may therefore have been different with the inclusion of a computerised measure of sustained attention.

Secondly, our BD sample size was modest: some have argued that CCA requires a sample size at least twenty times the number of variables in the analysis to adequately estimate the first canonical pair.[35] Others stated that a sample size of fifty is sufficient to detect strong canonical correlations (>.7).[42] Implementing PCA before CCA mitigates this issue to some extent,[34] but nevertheless, our results should be verified with a larger sample. Further, our sample size was too small to split into training and test datasets to perform an additional cross-validation step, which is often advised to validate the results of CCA.[34,35] While we utilised a large normative model of cortical thickness to generate robust estimations of cortical thickness abnormalities in BD, data were not available to do the same with neuropsychological scores. Therefore, estimations of cognitive impairment were made based only on the matched control group.

CCA is correlational, so we cannot infer cause-effect relationships. It may be the case the abnormal brain functioning leads to poorer cognitive function or vice-versa. Alternatively, both may be a consequence of disease burden. Other limitations of CCA include overfitting,[34] hence replication is important. Our first canonical correlation did not reach significance, therefore our interpretation of the brain-cognition associations detected here may not reflect true associations, or that our analysis was underpowered. Future studies should seek to replicate our results and may benefit from interpreting significant subsequent canonical pairs, which could provide distinct brain-cognition associations.[5]

Approximately half of our BD sample were taking lithium at the time of testing; the other half were taking maintenance treatments but were naïve to lithium. Studies suggest that, compared to patients not taking lithium, patients taking lithium have better EF and greater cortical thickness in several areas.[3,43] Lithium may therefore have a neuroprotective effect that may have influenced the results. We did not test the effects of lithium here because disaggregating the patient sample would have further reduced statistical power. Potential differences in cognitive or cortical data in lithium patients may not have been detrimental to our results since we assessed cortical thickness and cognitive impairments as continuous variables. However, it is possible that the nature of the continuous brain-cognition association might change in the presence of lithium, thus more research with larger samples is required.

## Heterogeneity and clinical confounds

An ongoing problem in researching the cognitive profile of BD is heterogeneity within the disorder: BD groups tend to vary in the degree of cognitive impairment, with some patients performing as well as healthy controls.[1] Patients with BD also vary in brain structural abnormalities.[44] Some evidence points towards the existence of subgroups with distinct cognitive profiles in BD, where the most impaired subgroup show unique brain morphology.[45] However, a recent review did not find strong evidence to support the idea that cognitive subgroups map onto distinct structural brain abnormality profiles in BD,[46,47] thus, the extent of heterogeneity in brain-cognition associations is not yet clear. Our BD sample appeared to vary in cognitive performance, but we accounted for the limitation of heterogeneity by considering cognitive functions and brain morphology as continuous variables that were quantified as a deviation from the healthy norm. However, the potential influence of cognitive heterogeneity should be considered in our results and in future studies.

Clinical confounds such as illness duration, illness severity, BD type (I or II), and psychosis may be related to cognitive impairment and brain abnormalities in BD and were not controlled for here.[1,3,36,48] Brain-cognition associations have been observed even after controlling for clinical confounds;[4] it may therefore be possible that the associations we found are independent of clinical characteristics, but further research should confirm this. Our sample was euthymic at the time of testing, so we could not assess the effect of mood state on brain-cognition associations. Other studies found that euthymic and depressed BD patients may differ in cortical thickness in some areas,[6] so future research should investigate the effect of mood on brain-cognition associations.

## Brain morphology

We used cortical thickness to measure structural abnormalities in BD. Previous research suggested that this may be a more sensitive measure of structural alterations, and more strongly related to cognitive impairment, in BD than other structural metrics.[3-5,46] However, cortical volume and surface area have also been associated with core cognitive functioning in BD, so should not be discounted in future studies.[11,49] We also acknowledge that we did not analyse white matter integrity or subcortical regions that are known to be involved in cognition in BD, such as the hippocampus, amygdala, and cerebellum.[50]

A limitation with measuring cortical thickness concerns mixed findings of whether poorer cognitive functioning is related to thinner or thicker cortex. Thinner cortex is generally assumed to be related to poorer cognitive functioning in BD,[6,48] however, some studies found that *increased* cortical thickness was related to cognitive impairment.[4,47] The direction of the relationship between cortical thickness and cognitive functioning depends on other features of cortical morphology: cortical thickness, volume, and surface area have been shown to covary in the healthy human brain according to a universal scaling law.[51] This may limit the interpretability of each metric alone. Wang et al. developed a sophisticated model of brain morphology that accounts for covariance between individual features and provides independent components that appear to be sensitive to structural differences in clinical groups.[51] Future research should utilise these independent components of cortical morphology to more precisely assess neural correlates of cognitive impairment in BD.

## Conclusion

This study is one of the few thus far to utilise multivariate methods to measure brain-cognition associations in BD and we have demonstrated the utility of CCA for this purpose. The results hint at associations between cortical thickness in temporal regions and impaired PS and EF in euthymic BD. Given the paucity of multivariate brain-cognition studies in BD, and difficulties with replicating brain-cognition associations, future research should use large datasets to further investigate the neural correlates of core cognitive functions in BD.

## CONTRIBUTIONS

The BLISS study was conceptualised and designed by DAC, AB, PT, and PG. AB and PT were responsible for the imaging acquisition at the Newcastle Magnetic Resonance Imaging Centre. CF was responsible for neuropsychological and clinical data collection, supervised by DAC. JPT and AB provided data for the normative model. This conceptualisation and design of this analysis and paper was done by BL, YW, and DAC. BL analysed the data and drafted the manuscript. All authors contributed to the final version of this manuscript.